\documentclass[twocolumn,balancelastpage,superscriptaddress,floatfix,longbibliography,nofootinbib,aps,pra,10pt]{revtex4-2}

\usepackage{stylesheet}
\usepackage{booktabs}
\usepackage{float}
\usepackage{xcolor}
\hypersetup{
    colorlinks,
    linkcolor={red},
    citecolor={blue},
}
\setcitestyle{square,numbers,comma}

\begin{document}

\notoc

\title{Average Rényi Entanglement Entropy in Gaussian Boson Sampling}

\date{May 7, 2025}

\author{Jason Youm}
\email{jasony2025@gmail.com}
\affiliation{\JQI}
\affiliation{\QUICS}
\affiliation{\MBHS}

\author{Joseph~T.~Iosue}
\email{jtiosue@umd.edu}
\affiliation{\JQI}
\affiliation{\QUICS}

\author{Adam~Ehrenberg}
\affiliation{\JQI}
\affiliation{\QUICS}

\author{Yu-Xin~Wang}
\affiliation{\QUICS}

\author{Alexey~V.~Gorshkov}
\affiliation{\JQI}
\affiliation{\QUICS}

\begin{abstract}
Recently, many experiments have been conducted with the goal of demonstrating a quantum advantage over classical computation. 
One popular framework for these experiments is Gaussian Boson Sampling, where quadratic photonic input states are interfered via a linear optical unitary and subsequently measured in the Fock basis.
In this work, we study the modal entanglement of the output states in this framework just before the measurement stage. 
Specifically, we compute Page curves as measured by various R\'enyi-$\alpha$ entropies, where the Page curve describes the entanglement between two partitioned groups of output modes averaged over all linear optical unitaries.
We derive these formulas for $\alpha = 1$ (\emph{i.e.}~the von Neumann entropy), and, more generally, for all positive integer $\alpha$, in the asymptotic limit of infinite number of modes and for input states that are composed of single-mode-squeezed-vacuum state with equal squeezing strength.
We then analyze the limiting behaviors when the squeezing is small and large.
Having determined the averages, we then explicitly calculate the Rényi-$\alpha$ variance for integers $\alpha > 1$, and we are able to show that these entropies are weakly typical.
\end{abstract}

\maketitle

\section{Introduction}
\label{sec:introduction}

Major advances in quantum computing over the past decade have opened up potential breakthroughs in various fields of science \cite{Möller2017,doi:10.1021/acs.chemrev.9b00829}. One such advancement is Gaussian Boson Sampling (GBS), an experimental framework for demonstrating quantum advantage that produces a sample from a distribution that is classically hard to compute \cite{PhysRevLett.113.100502, PhysRevLett.119.170501, PhysRevA.100.032326, doi:10.1126/sciadv.abl9236}. In this framework, squeezed photons are transformed by a Haar-random, passive Gaussian unitary consisting of beamsplitters and phase shifters. Although GBS is not computationally universal, it is nonetheless crucial for understanding the leverage of quantum devices over their classical counterparts at performing specific tasks.

The preparation of squeezed states is a much easier task than that of single photons used in Fock-state boson sampling, hence making GBS an experimentally favorable avenue toward demonstrating quantum advantage. Indeed, such demonstrations have already been realized in the lab \cite{zhong2020quantum-computa,zhong2021phase-programma,madsen2022quantum-computa}.

There is strong evidence that classically sampling from the output probability distribution of a GBS experiment is hard in the complexity-theoretic sense \cite{zhong2020quantum-computa, zhong2021phase-programma, madsen2022quantum-computa, preskill2012quantum, doi:10.1126/sciadv.abi7894}, whereas such sampling can be done efficiently on a quantum device. Moreover, entanglement is a critical asset across quantum computing, with additional applications in teleportation, dense coding, and quantum communication \cite{Yu_2021, wootters_quantum_1998, Zou_2021, RevModPhys.81.865}. Therefore, quantifying the behavior of entanglement in Gaussian bosonic systems has been of recent interest \cite{zhong2020quantum-computa, zhong2021phase-programma}, especially to sharpen the connection between entanglement and computational complexity.
Indeed,
the relationship between entanglement and complexity is generally unknown. 
Classical hardness requires at least some entanglement, but too much entanglement can also cause a state to be useless for computation. In fact, a randomly chosen finite-dimensional quantum state generally has too much entanglement to be computationally useful \cite{gross_most_2009,bremnerAreRandomPure2009}. 
This prompts our study of average and typical entanglement, and motivates further study into the general connection between entanglement and complexity.

In 1993, Page conjectured an explicit formula for the average entanglement entropy over all Haar-random pure states in an $N$-dimensional Hilbert space of the reduced density matrix on an $rN$-dimensional subsystem for $r\in [0, 1]$. This conjecture was proven soon after in the $N \to \infty$ limit, and the average entanglement entropy as a function of $r$ is now known as the \textit{Page curve} \cite{PhysRevLett.71.1291, PhysRevLett.72.1148, PhysRevE.52.5653}. The Page curve was initially studied in the context of information dynamics in and around black holes \cite{almheiri2021the-entropy-of-}, and has since found various applications in quantum information and condensed matter theory \cite{bianchi2022volume-law-enta}. As a natural follow-up, many works investigated the deviation of entanglement entropy from its expected value. A system exhibits \textit{typical entanglement} if the probability that a random state has entanglement bounded away from the average is small. This behavior, along with strong and weak classes of entanglement typicality, has been studied in black holes, thermodynamics, and other systems \cite{bianchi2022volume-law-enta,Hayden2006, Dahlsten_2014, PhysRevA.80.022317, Hayden2004, Popescu2006, PhysRevB.91.081110}.

Leaving the realm of standard finite-dimensional systems, the Page curve for fermionic Gaussian states has been extensively analyzed \cite{lydifmmode-zelse-zfiba2020eigenstate-enta,lydifmmode-zelse-zfiba2021entanglement-in,Bianchi_2021,bhattacharjee2021eigenstate-capa,ulifmmode-celse-cfiakar2022tight-binding-b,PhysRevResearch.5.013044, PhysRevD.106.046015}.
A natural next step, especially given the relevance to GBS, is to study bosonic Gaussian states. The typicality of entanglement in random bosonic Gaussian states was studied in Refs.~\cite{serafini2007teleportation-f,serafini2007canonical-and-m,fukuda2019typical-entangl,Iosue_2023}, and the Page curve for Gaussian bosonic systems was studied in Ref.~\cite{Iosue_2023}. However, Ref.~\cite{Iosue_2023} only computed the Page curve for the R\'enyi-2 entropy.

In this work, we derive analytic expressions for the R\'enyi-$\alpha$ Page curves of bosonic Gaussian systems for all $\alpha \in \bbZ_{>0}$, with special considerations for the von Neumann (Rényi-1) Page curve. The entropy of a mixed state is an indicator of its entanglement with the rest of the universe \cite{nielsen_quantum_2010}, and indeed the R\'enyi-$\alpha$ Page curve is a measure of the entanglement of an average state in a system as measured by the R\'enyi-$\alpha$ entropy, for $\alpha \in \mathbb{R}^{+}$.
The R\'enyi-$\alpha$ entropies offer a quantitative way to compare the amount of entanglement in two given quantum states, in that if the R\'enyi-$\alpha$ entanglement entropy of state $\ket{\psi_{1}}$ is greater than $\ket{\psi_{2}}$ for any given $\alpha$, then one cannot convert $\ket{\psi_{2}}$ into $\ket{\psi_{1}}$ via local operations and classical communication.
Each Rényi-$\alpha$ entropy has its own unique mathematical features, and notable examples are the von Neumann/Shannon entropy ($\alpha = 1$), collision entropy ($\alpha = 2$), and min-entropy ($\alpha \to \infty$). In particular, the min-entropy, the entropy defined by $-\log \text{max} ({p_i})$ for each $p_i$ a probability of occurrence of outcome $i$ when working with a discrete random variable, is a useful expression for working with extracting randomness from variables when lower entropies (e.g.~von Neumann) do not suffice \cite{5895072}. Therefore, it is natural to analyze the R\'enyi-$\alpha$ entropies in bosonic systems and extract their unique properties.

Our expressions are asymptotically exact in the number of modes $n$. Using these expressions, we prove various results about typicality of entanglement with respect to these Rényi-$\alpha$ entropies. 
A surprising feature that our analysis reveals is that R\'enyi-$\alpha$ entropies all concentrate for integer $\alpha$, which suggests that the entanglement spectrum of the states generated by Gaussian boson sampling also concentrate. This result could be useful in theoretical analysis of the performance of Gaussian boson sampling circuits.

Our setup is exactly that of a GBS experiment, and is the same as the setups considered in Refs.~\cite{fukuda2019typical-entangl, Iosue_2023}. A pure input quantum state with $n$ squeezed vacuum modes is acted on by a Haar-random, linear unitary (passive Gaussian) circuit. The Page curve, as a function of $r$, is the average R\'enyi-$\alpha$ entanglement between a group of $rn$ modes and the remaining system of $(1-r)n$ modes, where $r \in [0, 1].$ We expound our setup in greater detail in \cref{sec:setup}.

In \cref{sec:S1}, we present our results for the von Neumann entropy. Under an equal squeezing regime, we first present an explicit formula for the von Neumann Page curve in \cref{sssec:S1-formula} in terms of the partition ratio $r$ and initial squeezing strength $s$, which is exact up to terms that decay to zero as $n \rightarrow \infty.$ We subsequently discuss the von Neumann entropy under squeezing limits $s\rightarrow 0$ and $s\rightarrow \infty$ in \cref{sssec:S1-limits}.

We then generalize our analyses to the Rényi-$\alpha$ entropies for integers $\alpha \geq 2$
in \cref{sec:Salpha}. In \cref{sssec:Salpha-formula}, we derive the Rényi-$\alpha$ Page curve in terms of the partition ratio $r$ and initial squeezing strength $s$, with additional terms that vanish to zero as $n \rightarrow \infty$. We calculate squeezing limits of the Rényi-$\alpha$ entropy, this time with the extra dependence on $\alpha$, in \cref{sssec:Salpha-limits}. Finally, we evaluate the Rényi-$\alpha$ entropy under unequal, small initial squeezing strengths in \cref{sssec:Salpha-unequal} and derive typicality of entanglement results in \cref{sssec:Salpha-entanglement}.

\section{Setup}
\label{sec:setup}

In this section, we present our setup. We consider a bosonic system of $n$ modes. For a pedagogical introduction to the concepts discussed in the section, we refer to Ref.~\cite{serafini_quantum_2017}. For each $i \in \{1, 2, \dots, n\}$, define $r_{i} \coloneqq x_i$ to be the position operator and $r_{n+i} \coloneqq p_i$ to be the momentum operator. A density matrix $\rho$ describes a Gaussian state if there exists a $\beta > 0$ and a Hamiltonian $H$ that is at most quadratic in the quadrature operators $r_i$ such that $\rho$ is the thermal state $\rho \propto \e^{-\beta H}$.  When $\rho$ is Gaussian, it is fully characterized by its first and second moments $\Tr(\rho r_i)$ and $\sigma_{ij} = \frac{1}{2}\Tr[ \rho (r_i r_j + r_j r_i)] - \Tr(\rho  r_i)\Tr(\rho r_j)$. $\sigma$ is called the \textit{covariance matrix} of the state $\rho$.

In this work, we investigate the properties of bosonic states after the application of a linear optical unitary. Linear optical unitaries generate passive (boson-number conserving, or equivalently energy conserving) Gaussian unitaries. The set of passive Gaussian unitaries is isomorphic to $\text{Sp}(2n) \cap O(2n) \cong \U(n),$ where $\text{Sp}(2n)$ is the \textit{symplectic group} of $2n \times 2n$ matrices, $O(2n)$ is the \textit{orthogonal group} of $2n \times 2n$ matrices, and
$\U(n)$ is the \textit{unitary group} of $n\times n$ matrices \cite{hall2015lie-groups-lie-}.

We now define our notion of a random, pure Gaussian state on $n$ modes. We initialize the $i^{\text{th}}$ mode to be in a squeezed vacuum state with fixed squeezing parameter $s_i \in \bbR \, \, \forall \, i \in \set{1,\dots, n}$. We then randomly sample a passive Gaussian unitary $U \in \U(n)$ according to the Haar measure \cite{hall2015lie-groups-lie-} and apply it to the $n$ modes. The output state is characterized by the input squeezing strengths $s_i$ and the unitary $U$.
For squeezing parameters $s_1, s_2, \dots, s_n$, the total expected number of bosons in the state on the $n$ modes is $\sum_{i=1}^n \sinh^2(s_i)$. In this work, we primarily consider the case when all the squeezing parameters are equal, $s_i=s$. Then, the average total number of bosons per mode is $\sinh^2 (s)$.

We partition the $n$ output modes into two groups: one group of $k = r n$ modes for some $0\leq r \leq 1$, and the other group of $n-k = (1-r)n$ modes. We calculate the R\'enyi-$\alpha$ entropy of the reduced state of the $k$ modes. Because we are considering pure states, this is equivalent to the entropy of the reduced state of the $n-k$ modes \cite{nielsen_quantum_2010}. 

For the density matrix $\rho(U)$ on the $k$ modes, we denote its R\'enyi-$\alpha$ entropy by $S_\alpha(U) \coloneqq \frac{1}{1-\alpha} \log \Tr(\rho(U)^{\alpha})$. The limit as $\alpha \to 1$ yields the von Neumman entropy $S_1(U) \coloneqq -\Tr\rho(U)\log\rho(U)$.
We note that the dependence of these quantities on $r$, $n$, and $s$ is implicit. Furthermore, the average value of $S_\alpha(U)$ over $\U(n)$ with respect to the Haar measure is defined as the R\'enyi-$\alpha$ Page curve.

We will frequently utilize $\sigma(U)$, the covariance matrix of $\rho(U)$. We also use the \textit{symplectic eigenvalues} of $\sigma(U),$ defined as the eigenvalues of the matrix $i\Omega \sigma(U)$ where 
\begin{equation}
\label{eq:symplectic-form}
    \Omega \coloneqq 
    \begin{pmatrix} 
        0_{n \times n} & \mathbb{I}_{n \times n} \\ 
        -\mathbb{I}_{n \times n} & 0_{n \times n}
    \end{pmatrix}.
\end{equation}

\section{Von Neumann Entropy}
\label{sec:S1}

In this section, we calculate an explicit formula for the von Neumann Page curve as a function of equal squeezing strength $s$ and the partition size ratio $r$ up to terms that decay to zero as $n$ increases. Then, we compute and interpret the Page curve in the squeezing limits $s\rightarrow 0$ and $s\rightarrow \infty.$ Next, we discuss an approach to the case of small, unequal squeezing strengths. Finally, we comment on difficulties in deriving the variance and entanglement typicality of the von Neumann entropy in this setting.

\subsection{Explicit formula}
\label{sssec:S1-formula}

The von Neumann entropy is a function of the squeezing strength $s$ (assuming all modes are equally squeezed) and the partition size ratio $r = \frac{k}{n}$, up to terms that decay to zero as $n \rightarrow \infty.$ We explicitly calculate this function, which also leads us to the von Neumann Page curve as $n \rightarrow \infty.$

\medskip

\begin{theorem}[von Neumann entropy]
\label{thm:von-Neumann-Page-curve}

Choose any $s \in \mathbb{R}$ and $r \in [0, 1].$ Let $C_i = \frac{1}{i+1}\binom{2i}{i}$ represent the $i$th Catalan number and $_2F_1$ the hypergeometric function \cite{petkovsek1996, wolfram_hypergeometric2f1}. Then, the average von Neumann entropy over all unitaries $U \in U(n)$ is
\begin{widetext}
\begin{align}
    \begin{split}
        \underset{U \in U(n)}{\mathbb{E}} S_1(U) = \sum_{i=1}^{\infty}  \left[\frac{1}{2i} - \frac{1}{3}\textup{sech}^2 (2s) \tanh^{2i}(2s) \, _2F_1\left(\frac{3}{2}, 1+i, \frac{5}{2}, \textup{sech}^2(2s)\right)\right] 
        (nG_{i} (r) - H_{i}(r) + o(1)),
    \end{split}
\end{align}
\end{widetext}
where 
\begin{align}
    \label{eq:GH}
    G_{i}(r) &= r - r^{i+1}C_i \, _2F_1(1-i, i, 2+i, r), \\
    H_{i}(r) &= 4^{i-1} (r(1-r))^i
\end{align} are functions independent of $n$ detailed in Ref.~\cite{Iosue_2023}, and $o(1)$ denotes terms that vanish as $n\rightarrow \infty.$
\end{theorem}

The derivation for this formula is detailed in \cref{ap:S1-pagecurve}. In the derivation we use a formula for the von Neumann entropy in terms of the covariance matrix of the subsystem of $k$ modes. Using the decomposition of the covariance matrix derived in Ref.~\cite{Iosue_2023}, we Taylor expand about the identity matrix and utilize the negative binomial theorem. Finally, we use formulas from Ref.~\cite{Iosue_2023} regarding expectation values over $\U(n)$ of certain functions of matrix elements to arrive at the final expression.

The von Neumann Page curve as $n\rightarrow \infty$ is derived from \cref{thm:von-Neumann-Page-curve} by dividing by $n.$ The $H_{i}(r)$ and $o(1)$ terms on the RHS subsequently vanish. One important note is that the Page curve must be symmetric under $r \mapsto 1-r$ since the global state on the $n$ modes is pure \cite{nielsen_quantum_2010}. Since $G_i(r)$ and $H_i(r)$ are symmetric about $r = \frac{1}{2}$ \cite{Iosue_2023}, the von Neumann Page curve derived from \cref{thm:von-Neumann-Page-curve} is indeed symmetric about $r = \frac{1}{2}.$ This symmetry allows one to replace $r$ with $\min(r, 1-r).$

\subsection{Squeezing limits}
\label{sssec:S1-limits}

\cref{thm:von-Neumann-Page-curve} applies for all equal squeezing strengths $s$. We find considerable simplifications when examining the limiting behavior as a function of $s$.
We begin with the $s\to 0$ limit, which corresponds to photons in their near-vacuum states. One might expect the von Neumann Page curve to grow with the total number of bosons, $\sum_{i=1}^{n} \sinh s^2 \approx ns^2$ for small $s_i$. Indeed, this behavior occurs for the rest of the Rényi entropies (with integer $\alpha \geq 2$) as shown in \cref{sssec:Salpha-limits}.
However, the pertinent limit  $\lim_{s \rightarrow 0} \lim_{n \rightarrow \infty} \underset{U \in U(n)}{\mathbb{E}} \frac{1}{s^2n} S_1(U)$ diverges to infinity. In \cref{ap:S1-zerolimit}, we prove this result and demonstrate the relevant scaling to be $\underset{U}{\mathbb{E}}S_1(U) \sim s^2 \log(1/s^2)$, as formulated in the following theorem.

\begin{theorem}
For the squeezing strength limit $s\rightarrow 0$,
    \label{thm:von-Neumann-zero}
    \begin{align}
    \lim_{s \rightarrow 0} \lim_{n \rightarrow \infty} \underset{U \in U(n)}{\mathbb{E}} \frac{1}{\parentheses{s^2 \log(1/s^2)} n} S_1(U) &= r(1-r) .
\end{align}
\end{theorem}

We note that of course the entanglement must be zero when $s=0$ simply because there are no bosons present when $s=0$.
\cref{thm:von-Neumann-zero} addresses how the entanglement approaches zero as $s\to0$.
While the derivation of \cref{thm:von-Neumann-zero} is fairly straightforward, its significance is more interesting. The von Neumann Page curve's growth in order $s^2 \log(1/s^2)$ for small $s$ can be attributed to the unique definition of the von Neumann entropy $S_1(\rho) = - \Tr (\rho \log \rho),$ where $\rho$ is the density matrix of the quantum state. Therefore, the symplectic eigenvalues $\nu_j$ of the covariance matrix would also scale the von Neumann entropy in order $\nu_j \log \nu_j,$ warranting the $s^2 \log(1/s^2)$ dependence (as $\nu_j \sim s^2$). Furthermore, the von Neumann entropy is non-analytic unlike the rest of the Rényi entropies. Therefore, the argument presented in Ref.~\cite{Iosue_2023} that the Rényi entropy can be expanded in lowest order $s^2$ doesn't apply to the von Neumann case. More specifically, the deviation from analytic behavior arises from the von Neumman entropy formula in terms of the covariance matrix $\sigma$ \cite{wilde, serafini_quantum_2017, 10.21468/SciPostPhysCore.4.3.025}:
\begin{align}
    S_1 &= \frac{1}{2}\log \det\pargs{\frac{\sigma+\i \Omega}{2}} + \frac{1}{2}\Tr\pargs{\coth^{-1}(\i \Omega\sigma)\i\Omega\sigma},
\end{align}
where $\Omega$ is the symplectic form defined in \cref{eq:symplectic-form}. Since $\coth^{-1} (x)$ is not a real analytic function, the resulting Page curve is not analytic, either.

In a similar vein, we compute the von Neumann Page curve as $s \rightarrow \infty$, assuming an equal squeezing regime.

\begin{theorem}
In the strong squeezing limit $s\rightarrow \infty$,
    \label{thm:von-Neumann-inf}
    \begin{align}
    \lim_{s \rightarrow \infty} \lim_{n \rightarrow \infty} \underset{U \in U(n)}{\mathbb{E}} \frac{1}{sn} S_1(U) &= 2\min(r, 1-r).
    \end{align} 
\end{theorem}

Here, the von Neumann Page curve scales with $s$ as $s\rightarrow \infty.$ This is because, for large absolute squeezing values, higher orders in $r$ dominate the expansion of \cref{thm:von-Neumann-Page-curve}, which scale with $s.$ Interestingly, \cref{thm:von-Neumann-inf} remains the exact same for the Rényi-2 entropy \cite{Iosue_2023}, and indeed we will see in Sec.\ \ref{sssec:Salpha-limits} that it is the same for all the R\'enyi-$\alpha$ entropies for integral $\alpha$.

We now compare this result to the case of averaging over \textit{all} $n$-mode bosonic states with a fixed boson number constraint. 
In particular, 
the volume (\textit{i.e.}~proportional to $n$) scaling of the corresponding Page curve is \cite{yauk2024typical-entangl}
\begin{equation}
\begin{split}
\lim_{n\to\infty}&\frac{1}{n} S_1^{(n_b)} = \\
&\parentheses{(r_b + 1) \log(1 + r_b) - r_b \log(r_b)}
\min(r,1-r),
\end{split}
\end{equation}
where $r_b n = n_b$ the number of bosons.
In this case, the Hilbert space is finite-dimensional with dimension $\binom{n+n_b-1}{n_b}$.
As $n_b\to\infty$, this yields 
$\frac{1}{n}S_1^{(n_b)} \sim \log(r_b)\min(r,1-r) $

In our Gaussian setup, the expected number of bosons is $\angles{n_b} = n \sinh^2(s)$.
In this case, the Hilbert space in infinite-dimensional, but the fixed squeezing constraint gives an energy constraint.
Using \cref{thm:von-Neumann-inf}, we then see that as $n_b\to\infty$ ,
$\frac{1}{n}S_1 \sim \log\pargs{\angles{r_b}} \min(r,1-r)$.

We therefore see that, asymptotically, the average entanglement scales proportional to the volume of the system. Given a fixed but asymptotically large energy (\textit{i.e.~}number of bosons), the average entanglement over \textit{Gaussian} states equals the average entanglement over \textit{all} bosonic states.

Finally, we also note that when $r\to 0$, all of the Page curves that we have computed, and indeed all of the Page curves computed in Ref.~\cite{yauk2024typical-entangl},
scale proportionally to $r$.
In particular, we can see from 
Ref.~\cite[Fig.~2, Eq.~(23)]{yauk2024typical-entangl} that the fixed particle number Page curves for (a) fermions, (b) hardcore bosons, and (c) (multi-species) bosons all have Page curves that are proportional to 
$r$ as $r\to0$.
Similarly, Page curves in finite-dimensional spaces without constraints are also proportional to $r$ as $r\to 0$ \cite{PhysRevLett.71.1291, PhysRevLett.72.1148, PhysRevE.52.5653,bianchi2022volume-law-enta,Hayden2006, Dahlsten_2014, PhysRevA.80.022317, Hayden2004, Popescu2006, PhysRevB.91.081110}.
In the Gaussian case that we have computed in this work, the Page curve is also proportional to $r$ in this limit, as can be seen in \cref{thm:von-Neumann-Page-curve,thm:Renyi-Page-curve}.

\subsection{Unequal squeezing}
\label{sssec:S1-unequal}

In Ref.\ \cite{Iosue_2023}, the expected value of the Rényi-2 entropy for small, unequal squeezing strengths was calculated through a power-series analysis. The analyticity of the Rényi-2 function was what allowed writing the entropy as a power series in $s_i,$ where $s_i$ was the squeezing strength of the $i$th mode. Using this power series expansion, along with the translational invariance of the Haar measure, Ref.\ \cite{Iosue_2023} derived a formula for unequal squeezing in order $s^2,$ which reduces to the $s\rightarrow 0$ squeezing limit for small, equal $s_i.$

However, due to the von Neumann entropy not being analytic around $s=0,$ we cannot perform the same analysis to derive a formula for unequal squeezing strengths. Furthermore, directly approaching the unequal squeezing regime via the Taylor expansion of the von Neumann entropy runs into another problem, since the separation between $s$ and $r$ dependence present in the Rényi-$2$ case presented in Ref.~\cite{Iosue_2023} no longer applies here. Specifically, the matrix $W$ described in \cref{ap:S1-pagecurve} is no longer independent of $s.$ Therefore, carrying out the Taylor expansion would yield complicated matrix expressions in $\sigma$ that do not simplify appreciably at first glance.
Due to these difficulties, we are unable to derive the unequal squeezing case for the von Neumann entropy. It is nonetheless possible, though, that executing the complicated matrix Taylor expansion and simplifying terms yields a closed-form result for the von Neumann entropy.

\medskip

\section{General \texorpdfstring{Rényi-$\alpha$}{Rényi-alpha} Entropies}
\label{sec:Salpha}

\medskip

\begin{figure*}
    \centering
    \includegraphics[width=.9\textwidth]{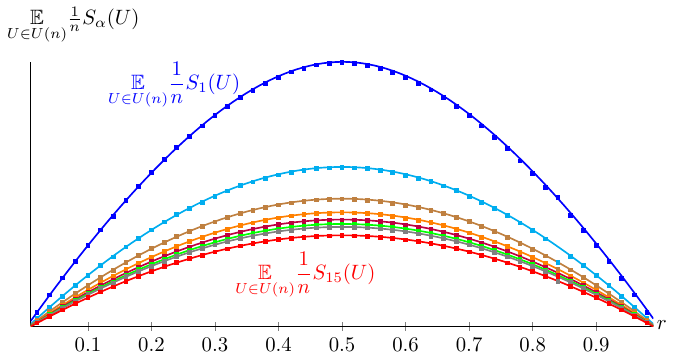}
    \caption{Simulated Page curves for Rényi-$\alpha$ entropies for $\alpha \in \{1, 2, 3, 4, 5, 6, 7, 15\}$ in dots (smaller Page curves indicate larger $\alpha$), overlaid on the analytic expressions derived in \cref{thm:von-Neumann-Page-curve} and \cref{thm:Renyi-Page-curve}. The simulated values were run 250 times. The number of modes is $n=400$ with equal squeezing strength $s=0.5$.}
    
    \label{fig:Renyi-Page-curves}
\end{figure*}

In this section, we extend much of our analysis with the von Neumann entropy in \cref{sec:S1} to Rényi-$\alpha$ entropies for integer $\alpha \geq 2$. We also derive results for the typicality of entanglement and the small, unequal squeezing case. First, we present the Rényi-$\alpha$ Page curve $\forall \alpha \in \mathbb{Z}, \alpha \geq 2.$ Next, we investigate squeezing limits as $s\rightarrow 0$ and $s\rightarrow \infty,$ as well as consider the Rényi-$\alpha$ Page curve for small, unequal squeezing strengths.
We also study the variance of the entanglement and prove various regimes of entanglement typicality. Finally, we show some numerical simulations confirming our results.

\subsection{Explicit formula}
\label{sssec:Salpha-formula}
As before, we consider a system of $n$ modes, each with equal initial squeezing strength $s.$

\begin{theorem}[Rényi-$\alpha$ Page curve]
For all integer $\alpha \geq 2,$ define $a = \floor{\frac{\alpha - 1}{2}}$. Let $\zeta = 1$ if $\alpha$ is even and $\zeta = 0$ if $\alpha$ is odd.
Then, the Rényi-$\alpha$ entropy is given by
\label{thm:Renyi-Page-curve}

\begin{widetext}
\begin{align}
    \begin{split}
        \underset{U \in U(n)}{\mathbb{E}} S_\alpha(U) = \frac{ \zeta }  {\alpha -1 } ~ \underset{U \in U(n)}{\mathbb{E}} S_2
        + \frac{1}{\alpha -1 }
        \sum _{i=1}^{\infty } \left [
        \sum _{m=1} ^{ a }
        \frac{ \sinh^{2i}(2s) }
            {
        i (\cosh^2(2s) + \cot ^{2}
        \frac{\pi m }{\alpha} )^{i}}  
        \right ]
        \left(nG_{i} (r) - H_{i} (r) + o(1)\right)
        ,
    \end{split}
\end{align}
\end{widetext}
where $ \underset{U \in U(n)}{\mathbb{E}} S_2$ is the expected value of the Rényi-2 entropy as a function of $s$ and $r$ given as
\begin{align}
     \underset{U \in U(n)}{\mathbb{E}} S_2 = \sum_{i=1}^{\infty} \frac{\tanh^{2i}(2s)}{2i} \parentheses{n G_{i}(r) - H_i(r) + o(1) },
\end{align}
and where $G_i (r)$ and $H_i (r)$ are defined in \cref{eq:GH}.

\end{theorem}
The proof of \cref{thm:Renyi-Page-curve} is shown in \cref{ap:Salpha-pagecurve}. In the proof, we use a formula for the R\'enyi-$\alpha$ entropy in terms of the symplectic eigenvalues of $\sigma,$ the covariance matrix of the reduced state of $k$ modes (see, e.g., Ref.~\cite{serafini_quantum_2017}). We convert this formula into a matrix expression in terms of $\Omega$ and $\sigma$, and perform significant simplifications. Finally, to compute the expectation value over the unitary group, we again utilize averaged matrix quantities derived in Ref.~\cite{Iosue_2023}.

Since $G_i(r)$ is symmetric under $r\to 1-r$, we see that the Page curves are also manifestly symmetric under $r\to 1-r$. Various analytic Rényi-$\alpha$ Page curves are plotted in \cref{fig:Renyi-Page-curves}. We also plot numerically-simulated values of the Page curve, where we generate Haar-random unitary matrices and compute the entropies using the positive symplectic eigenvalues $\nu_j$ of each covariance matrix $\sigma$. Then, we compute the respective Rényi-$\alpha$ entropy as $S_{\alpha}(U) = \sum_{j=1}^{n} \frac{1}{1-\alpha} \ln\left(\frac{2^{\alpha}}{(\nu_j+1)^{\alpha}-(\nu_j-1)^{\alpha}}\right)$ \cite{serafini_quantum_2017}. The numerical and analytical values overlap, indicating the validity of \cref{thm:von-Neumann-Page-curve} and \cref{thm:Renyi-Page-curve}.

\subsection{Squeezing limits}
\label{sssec:Salpha-limits}

We now investigate the Rényi-$\alpha$ Page curve for all integer $\alpha \geq 2$ in the squeezing limits $s\rightarrow 0$ and $s\rightarrow \infty.$ We find that, just as in the Rényi-2 case, the Page curve scales with $s^2$ when $s$ approaches zero and scales with $s$ when $s\to\infty$. 

\begin{theorem}
    For integer $\alpha \geq 2,$ the Page curves of the Rényi-$\alpha$ entropy as the equal squeezing strength $s$ approaches $0$ and $\infty,$ respectively, are given by
    \label{thm:Renyi-zero-inf}
    \begin{align}
    \lim_{s \rightarrow 0} \lim_{n \rightarrow \infty} \underset{U \in U(n)}{\mathbb{E}} \frac{1}{n s^2} S_{\alpha}(U) &= \frac{\alpha}{\alpha - 1} r(1-r) , \\
    \lim_{s \rightarrow \infty} \lim_{n \rightarrow \infty} \underset{U \in U(n)}{\mathbb{E}} \frac{1}{ns} S_{\alpha}(U) &= 2 \min (r, 1-r) .
    \end{align}
\end{theorem}

In general, the Rényi entropies satisfy $S_\alpha \geq S_{\alpha'}$ whenever $\alpha \leq \alpha'$. This behavior matches with our formulas for the Rényi-$\alpha$ limits in \cref{thm:Renyi-zero-inf}, as the Page curve decreases when $\alpha$ increases. From \cref{thm:Renyi-zero-inf}, it is furthermore evident that $S_2(U) = 2S_{\infty}(U)$ when $s\to 0$.
This relation between two Rényi entropies saturates the case of a discrete random variable with outcome probabilities $p_1, p_2, \dots, p_m,$ where
\begin{align}
    S_2 = -\log \sum_i p_i^2 \leq -\log \sup p_i^2 = -2 \log \sup p_i = 2 S_{\infty}.
\end{align}

In the infinite squeezing case, the Page curve remains independent of $\alpha$ in order $s$ as $\alpha \rightarrow \infty$. This indicates that, as $s \to \infty,$ the Page curve approaches the curve of
maximum Gaussian entanglement (that is, the maximum is taken over all Gaussian states with a fixed initial squeezing configuration).

\medskip

\setlength{\tabcolsep}{12pt}

\begin{table*}[ht]
    \centering
    \begin{tabular}{ccccc}
         \toprule
         & & $k = \bigTheta{n}$ & $k = \littleo{n}$ \cite{Iosue_2023} & $k = \littleo*{n^{1/3}}$ \cite{fukuda2019typical-entangl} \\
        \midrule
         \textbf{Equal} & \textit{Rényi-$\texorpdfstring{\alpha}{alpha}$ ($\alpha \geq 2$)} & weak & strong & strong\\
         \textbf{squeezing} & \textit{von Neumann} & ? & weak & strong \\\\
         \textbf{Unequal} & \textit{Rényi-$\texorpdfstring{\alpha}{alpha}$ ($\alpha \geq 2$)} & ? & weak${}^\ast$ & strong \\
         \textbf{squeezing} & \textit{von Neumann} & ? & weak${}^\ast$ & strong \\
        \bottomrule
    \end{tabular}
    \caption{
    A summary of the current status of rigorous results on entanglement typicality in Gaussian bosonic systems. This table is the same as the one provided in Ref.~\cite{Iosue_2023}, but the Rényi-2 case has been generalized for all Rényi-$\alpha$ entropies where $\alpha \in \mathbb{Z}^{+}.$ Note that ``weak${}^\ast$'' indicates that the result is not fully proven, but is based on Conjecture 10 in Ref.~\cite{Iosue_2023}. Furthermore, wherever "weak" typicality is displayed, we have not ruled out the possibilty of strong typicality. 
    We present typicality results when $k,$ the number of modes in the subsystem, scales with $\bigTheta{n}, \littleo{n},$ and $\littleo{n^{1/3}},$ respectively. Note that we only take the "worst-case" typicality; if there exists both weak and strong typicality for one scaling, we display only the weakly typical case. We present typicality results separately for the Rényi-$\alpha$ entropies and the von Neumann entropy. The leftmost column ($k = \bigTheta{n}$) stems from our findings in \cref{sssec:Salpha-entanglement}, while the middle column ($k = \littleo{n}$) follows from Ref.~\cite{Iosue_2023}. The rightmost column ($k = \littleo{n^{1/3}}$) follows from the results of Ref.~\cite{fukuda2019typical-entangl}. Refs.~\cite{serafini2007teleportation-f, serafini2007canonical-and-m} proved strong typicality in the regime $k = \bigO{1}$.
    }
    \label{tab:typicality}
\end{table*}

\subsection{Unequal squeezing}
\label{sssec:Salpha-unequal}

We now consider the Rényi-$\alpha$ entropy when the initial squeezing strengths are not necessarily equal, but are close to zero. We let the squeezing strength on the $i$th mode be $s_i$ for some $s_i \in \mathbb{R}.$ Define $s_{\text{max}} \coloneqq \max{(|s_1|, |s_2|, \dots, |s_n|})$.

\begin{theorem}
    For integer $\alpha \geq 2,$ the Rényi-$\alpha$ entropy for small, unequal squeezing strengths $s_1, s_2, \dots, s_n$ is given by
    \label{thm:Renyi-unequal}
    \begin{align}
    \Expval_{U\in\U(n)}S_{\alpha}(U) = \frac{\alpha}{\alpha - 1} r(1-r)\sum_{i=1}^n s_i^2 + \bigO{rn s_{\rm max}^4}.
    \end{align}
\end{theorem}

To prove \cref{thm:Renyi-unequal}, note that the Rényi-$\alpha$ entropy is real analytic for positive integer $\alpha$. This is because the Rényi-$\alpha$ expression, when written in terms of the covariance matrix $\sigma$, contains only analytic expressions in the vicinity of $s=0$, and thus the final expression is also analytic in the small-squeezing limit \cite{serafini_quantum_2017}. Thus, we can write the entropy as a power series in $s_i$. Furthermore, there exists a passive, local (i.e.\ acting independently on each mode) Gaussian unitary that acts on the initial covariance matrix $\sigma_0$ with the transformation $s_i \rightarrow -s_i$. Since this unitary can absorbed into the Haar random unitary $U$, the power series can be expanded in $s_i^2$. Note that when the absolute squeezing strengths are close to zero, the first statement of \cref{thm:Renyi-zero-inf} is a special case of \cref{thm:Renyi-unequal}.

There exists another passive Gaussian unitary that acts on $\sigma_0$ via the transformation $s_i \to s_{\tau(i)}$ for any permutation $\tau$ of $n$ elements. Since this unitary can also be absorbed into the Haar random unitary $U$, the power series is symmetric under $s_i \to s_{\tau(i)}$. Thus, the $s_i^2$ term in the power series expansion is of the form $g(r, n) \sum_{i=1}^{n} s_i^2$ for some function $g(r, n)$. The equal-squeezing result in \cref{thm:Renyi-zero-inf} then immediately implies $g(r, n) = \frac{\alpha}{\alpha-1} r(1-r).$

\subsection{Typicality of entanglement}
\label{sssec:Salpha-entanglement}

Next, we investigate the variance of the Rényi-$\alpha$ Page curve. Chebyshev's inequality allows one to relate the variance with the \textit{typicality of entanglement}, where typicality describes how quickly the difference between some quantity (in this case, entanglement) and its average vanishes. This analysis is useful in statistical mechanics, where typicality is useful for thermodynamic variables to accurately represent their mean values.

The formal definition of typicality as described in Ref.~\cite{Iosue_2023} is as follows: a unitary-dependent nonnegative random variable $S(U)$ is $\textit{weakly typical}$ if, for any constant $\epsilon > 0$,
\begin{align}
    \lim_{n\rightarrow \infty} \underset{U\in U(n)}{\text{Pr}} \left[\left|\frac{S(U)}{\underset{V \in U(n)}{\mathbb{E}} S(V)} - 1\right| < \epsilon \right] = 1.
\end{align}
Similarly, $S(U)$ is $\textit{strongly typical}$ if, for any constant $\epsilon > 0$,
\begin{align}
    \lim_{n\rightarrow \infty} \underset{U\in U(n)}{\text{Pr}} \left[\left|S(U) - \underset{V \in U(n)}{\mathbb{E}} S(V) \right| < \epsilon \right] = 1.
\end{align}

Perhaps contrary to intuition, a weakly typical variable has its multiplicative distance vanish in the thermodynamical limit, while a strongly typical variable has its additive distance vanish. This occurs because the entropy goes to infinity as $n \rightarrow \infty.$ In this regard, strong typicality implies weak typicality.

We are able to prove typicality results based on a crucial invariance described in Ref.~\cite{Iosue_2023}.
Specifically,  Ref.~\cite{Iosue_2023} derives that the variance of the Rényi-2 entropy is constant as $n \to \infty$. We extend this to general integer $\alpha \geq 2$ using a Cauchy-Schwarz argument in \cref{ap:Salpha-entanglement}. Using a bound for the variance via Chebyshev's inequality, the Rényi-$\alpha$ entropy exhibits weak typicality for $k = \Theta(n)$ (again, for all integer $\alpha \geq 2$). We present our results, as well as past analyses of entanglement typicality in Gaussian bosonic systems, in \cref{tab:typicality}.

\section{Conclusions and open questions}
\label{sec:conclusion}

In this paper, we have carefully studied the Rényi and von Neumann entropies of random bosonic Gaussian states. Such states are exactly the output states of a GBS experiment. Specifically, we have found closed formulas for the Rényi-$\alpha$ entanglement entropy in terms of the squeezing strength $s$ and the mode partition ratio $r$ that are exact asymptotically in $n$, where $\alpha \in \mathbb{Z}^{+}.$ Using this, we calculated the Rényi-$\alpha$ Page curve and various results stemming from it, such as $s\rightarrow 0$ and $s\rightarrow \infty$ squeezing limits, and, for $\alpha \geq 2$, variance and typicality of entanglement.

We have not reached a conclusion regarding the typicality of entanglement for the von Neumann entropy as $k$ scales with $\Theta(n).$ We have that $\variance_U S_1(U) = \Expval_U S_1(U)^2 - \parentheses{\Expval_U S_1(U)}^2,$ which we have not been able to simplify appreciably when using the von Neumann entropy formula from \cref{thm:von-Neumann-Page-curve}. Specifically, the squaring of hypergeometric functions yields terms in the variance expansion that we have not been able to simplify. More research should be done to determine the typicality of entanglement for the von Neumann entropy. However, since we've proven in \cref{ap:Salpha-entanglement} that the Rényi-$\alpha$ entropies for integer $\alpha \geq 2$ are only weakly typical, the von Neumann entropy is unlikely to be strongly typical. It remains to be rigorously proven, perhaps through examining special cases of the von Neumann entropy, such as when $r = \frac{1}{2}$ or $s \to 0,$ and attempting to extract the variance from these cases.

Another point of research is in generalizing the Rényi-$\alpha$ entropy in \cref{thm:Renyi-Page-curve} to non-integer $\alpha$. In our work, we only considered $\alpha \in \mathbb{Z}^{+}.$ Extending to non-integer $\alpha$ would allow a formula for the Rényi entropies that is continuous in $\alpha$, and could be useful to examine the behavior of the Rényi entropy as $\alpha \rightarrow 1$.

We also propose other areas of further research. Thus far, we have only considered Haar-random unitaries that describe the linear optical circuit. However, an interesting point of research could be in examining the entanglement associated with random linear optical circuits of finite depth, that is, the \textit{depth dependence} of entanglement entropy. Currently, the precise relationship between entanglement and  
circuit depth is largely unknown \cite{Brod_2015}. There have been some numerical studies \cite{PhysRevA.99.062334}, but only specialized cases where most, but not all, initial modes are vacuum are understood analytically \cite{Zhang2021}.

Another related notion deals with \textit{sampling complexity}, the overall complexity of classical computers producing a sample from approximately the same distribution as the quantum system 
\cite{Hangleiter_2023}. Quantifying the relationship between entanglement and complexity, and studying the depth dependence of entanglement and complexity, could yield crucial new insights into the classical hardness of simulating quantum experiments such as GBS and, more generally, into the relation between entanglement and complexity.

Another problem to tackle deals with unequal squeezing. There are two primary avenues of future research. First, for small, varying squeezing strengths $s_1, s_2, \dots, s_n,$ what is the average von Neumann entanglement entropy of the subsystem? We have outlined a method in \cref{sssec:S1-unequal} of tackling this problem, but initial attempts have proved futile due to the heavy presence of inverse matrices. Nonetheless, learning the behavior of the von Neumann entropy for unequal squeezing regimes is important from a practical standpoint, since many GBS experiments are conducted with one set of modes at zero squeezing strength and the other at some varied squeezing strength close to zero. Second, it is interesting to consider Rényi-$\alpha$ entropies for systems with varying squeezing strengths that are not necessarily close to zero. The argument given in \cref{sssec:Salpha-unequal} revolves around the entropy expression being analytic around $s=0$ and vanishing in higher orders of $s_{\text{max}}.$ However, for the most general case of $s_{\text{max}} \in \mathbb{R},$ one cannot simply extend the $s\rightarrow 0$ squeezing limit for unequal squeezing. Nonetheless, deriving such an expression would be important for a wider range of bosonic systems.

\begin{acknowledgments}
    This work was supported in part by the U.S.~Department of Energy, Office of Science, Accelerated Research in Quantum Computing, Fundamental Algorithmic Research toward Quantum Utility (FAR-Qu). Additional support is acknowledged from DARPA SAVaNT ADVENT, AFOSR MURI, DoE ASCR Quantum Testbed Pathfinder program (awards No. DE-SC0019040 and No. DE-SC0024220), NSF QLCI (award No. OMA-2120757), NSF STAQ program, AFOSR, ARL (W911NF-24-2-0107), and NQVL:QSTD:Pilot:FTL
    Support is also acknowledged from the U.S.~Department of Energy, Office of Science, National Quantum Information Science Research Centers, Quantum Systems Accelerator.  
    JTI thanks the Joint Quantum Institute at the University of Maryland for support through a JQI fellowship.
    Y.-X.W.~acknowledges support from a QuICS Hartree Postdoctoral Fellowship.
    Specific product citations are for the purpose of clarification only, and are not an endorsement by the authors or NIST.
\end{acknowledgments}

\toc

\clearpage
\onecolumngrid\appendix

\section{von Neumann Entropy}
\label{ap:S1-appendix}

In this section of the appendix, we show derivations of our results regarding the von Neumann entropy. We cover the derivation of the von Neumann entropy presented in \cref{thm:von-Neumann-Page-curve}, show squeezing limits as $s\rightarrow 0$ and $s\rightarrow \infty,$ (\cref{thm:von-Neumann-zero} and \cref{thm:von-Neumann-inf}), and discuss a potential avenue for solving the unequal squeezing case.

\subsection{Page Curve for the von Neumann Entropy}
\label{ap:S1-pagecurve}
Here, we derive the von Neumann entanglement entropy between the subsystems with $k$ and $n-k$ modes, in terms of the partition size ratio $r = \frac{k}{n}$ and the squeezing strength $s.$ We use much of the notation from Ref.~\cite{Iosue_2023}, and the background on bosonic Gaussian states and operations can be found in Ref.~\cite{serafini_quantum_2017}.

Consider a system of $n$ bosonic modes, where each mode $1\leq i\leq n$ is prepared in a single-mode squeezed state with squeezing strength $s_i.$ This $n$-mode Gaussian state is acted on by a passive linear optical unitary $U \in U(n).$

We divide the output modes into two groups, with one group bearing $0\leq k\leq n$ modes. Let $\sigma$ be the $2k \times 2k$ covariance matrix of the reduced state on these $k$ modes. As shown in Refs.~\cite{serafini_quantum_2017, wilde, 10.21468/SciPostPhysCore.4.3.025}, the von Neumann entanglement entropy $S_1(U)$ of the whole system is given by
\begin{align}
    \label{eq:A2}
    S_1(U) &= \frac{1}{2}S_1^{(1)} + \frac{1}{2} S_1^{(2)} \, \, \text{ with } \, \,
    \begin{cases} 
      S_1^{(1)} = \log \det\pargs{\frac{\sigma+\i \Omega}{2}} \\
      S_1^{(2)} = \Tr\pargs{\coth^{-1}(\i \Omega\sigma)\i\Omega\sigma}
   \end{cases} ,
\end{align}
where $\coth^{-1}(x)$ is the inverse hyperbolic cotangent function, and the symplectic form $\Omega$ equals
\begin{align}
    \Omega := \begin{pmatrix} 0_{n \times n} & \mathbb{I}_{n \times n} \\ -\mathbb{I}_{n \times n} & 0_{n \times n} \end{pmatrix}.
\end{align}
We aim to derive a formula for $S_1$ solely in terms of $r, n, $ and $s$ using \cref{eq:A2}.

Define the $k \times n$ matrix $P$ and the $n\times n$ projector $\Pi$ such that
\begin{align}
    P &:= \begin{pmatrix} \mathbb{I}_{k \times k} & 0_{k \times (n-k)} \end{pmatrix}, \\
    \Pi &:= \begin{pmatrix} \mathbb{I}_{k \times k} & 0_{k \times (n-k)} \\ 0_{(n-k) \times k} & 0_{(n-k) \times (n-k)}\end{pmatrix} = \mathbb{I}_{k \times k} \oplus 0_{(n-k) \times (n-k)}.
\end{align}
From Ref.~\cite{Iosue_2023}, when $s_i = s_j \, \, \forall i, j \in \set{1, 2, \dots, n},$ the covariance matrix $\sigma$ of the reduced state after the application of $U$ can be expressed as
\begin{align}
    \label{eq:A5}
    \sigma = \cosh{(2s)}\mathbb{I} + \sinh{(2s)}M, \, \, \text{ where } \, \, M := \begin{pmatrix} P\, \text{Re}(\bar{U}U^{\dag}) \, P^{T} & P\, \text{Im}(\bar{U}U^{\dag}) \, P^{T} \\ P\, \text{Im}(\bar{U}U^{\dag}) \, P^{T} & -P\, \text{Re}(\bar{U}U^{\dag}) \, P^{T} \end{pmatrix}
    ,
\end{align}
and $\bar{U}, U^{T}, U^{\dag}$ represent the conjugate, transpose, and conjugate transpose of $U,$ respectively. Two useful and easily checkable properties of $M$ are that $\Omega M = -M \Omega$ and $\text{Tr}(\Omega M^{j}) = \text{Tr}(M^{2j-1}) = 0, \, \, \forall j \in \mathbb{Z}^{+}.$ 

Using \cref{eq:A5} and the properties above, $S_1^{(1)}$ from \cref{eq:A2} is expressed as
\begin{align}
    S_1^{(1)}
    &= \Tr \log\pargs{\frac{\sigma+\i\Omega}{2}} \\
    &= \sum_{j=1}^\infty \frac{(-1)^{j+1}}{j} \Tr \parentheses{\frac{\sigma+\i\Omega}{2} - \bbI}^j 
    \nonumber \\
    &= \sum_{j=1}^\infty \frac{(-1)^{j+1}}{j 2^j} \Tr \parentheses{(\cosh(2s)-2)\bbI + \sinh(2s) M+\i\Omega}^j \nonumber \\
    &= \Tr \sum_{j=1}^\infty \frac{(-1)^{j+1}}{j 2^j} 
    \sum_{\ell=0}^j \binom{j}{\ell} \parentheses{\sinh(2s) M+\i\Omega}^\ell  (\cosh(2s)-2)^{j-\ell} \nonumber \\
    &= \sum_{j=1}^\infty \frac{(-1)^{j+1}}{j 2^j} \parentheses{
    2k (\cosh(2s)-2)^j +
    \Tr \sum_{\ell=1}^j \binom{j}{\ell} \parentheses{\sinh(2s) M+\i\Omega}^\ell  (\cosh(2s)-2)^{j-\ell}} \nonumber \\
    &= 2k \log\left(\frac{\cosh 2s}{2}\right)-\Tr \sum_{\ell=1}^\infty \parentheses{\sinh(2s) M+\i\Omega}^\ell \sum_{j=\ell}^\infty \frac{(-1)^j}{j 2^j}
     \binom{j}{\ell}   (\cosh(2s)-2)^{j-\ell} \nonumber \\
    &= 2k \log\left(\frac{\cosh 2s}{2}\right)-\Tr \sum_{\ell=1}^\infty \frac{(-1)^{\ell}}{\ell} \parentheses{\sinh(2s) M+\i\Omega}^\ell \cosh^{-\ell}(2 s) \nonumber \\
    &= 2k \log\left(\frac{\cosh 2s}{2}\right)-\Tr \sum_{\ell=1}^\infty \frac{1}{2\ell} \parentheses{\sinh(2s) M+\i\Omega}^{2\ell} \cosh^{-2\ell}(2 s) \nonumber \\
    &= 
    \label{eq:A14}
    2k \log\left(\frac{\cosh 2s}{2}\right)-\Tr \sum_{\ell=1}^\infty \frac{1}{2\ell} \cosh^{-2\ell}(2 s)
    \parentheses{\sinh^2(2s)M^2+\bbI}^\ell.
\end{align}

As discussed in Ref.~\cite{Iosue_2023}, let $W = \Pi U \bar{U}^{\dag}\Pi \bar{U} U^{\dag}\Pi.$ Then, $\text{Tr}(M^{2i}) = 2 \, \text{Tr}(W^i)$ $\forall i\in \mathbb{Z}^{+}.$ We emphasize that the dependence on $U$ and $r$ is implicit in $W$. Importantly, though, there is no $s$ dependence in $W$. We also note that, since now everything is written in terms of traces, we are able to replace $P$ with $\Pi$ so that all matrices are square. Incorporating this definition into \cref{eq:A14}, we have
\begin{align}
    S_1^{(1)}
    &= 2k \log\left(\frac{\cosh 2s}{2}\right)- 2\sum_{\ell=1}^\infty \frac{1}{2\ell} \cosh^{-2\ell}(2 s) \sum_{j=0}^\ell \binom{\ell}{j} \sinh^{2j}(2s) \Tr W^j \\
    &= 2k \log\left(\frac{\cosh 2s}{2}\right)- 2\sum_{\ell=1}^\infty \frac{1}{2\ell} \cosh^{-2\ell}(2 s) \parentheses{
    k
    +
    \sum_{j=1}^\ell \binom{\ell}{j} \sinh^{2j}(2s) \Tr W^j
    } \nonumber \\
    &= 2k \log\left(\frac{\cosh 2s}{2}\right) + k \log \bargs{\tanh^2 (2s)} - 2\sum_{j=1}^\infty \sinh^{2j}(2s)\Tr W^j \sum_{\ell=j}^\infty \frac{1}{2\ell \cosh^{2\ell}(2s)} \binom{\ell}{j} \nonumber \\
    &= k \log \bargs{\cosh^2 (s) \sinh^2 (s)} - 2\sum_{j=1}^\infty \sinh^{2j}(2s)\Tr W^j \frac{1}{2j\sinh(2s)^{2j}} \nonumber  \\
    &= k \log \bargs{\frac{\sinh^2 (2s)}{4}} - \sum_{j=1}^\infty \frac{1}{j} \Tr W^j .
\end{align}

We use a similar approach to simplify $S_1^{(2)}.$ Notably, we use the Taylor expansion 
\begin{align}
    \coth^{-1}(x) = \sum_{j=0}^{\infty} \frac{x^{-2j-1}}{2j+1} \, \, \text{ for } \, \, |x| \geq 1.
\end{align}
The $|x| \geq 1$ condition is met, as the eigenvalues of $i \Omega \sigma$ come in positive-negative pairs, each with magnitude greater than (or equal to) 1. Thus,
\begin{align}
    S_1^{(2)} 
    &= \Tr (\coth^{-1}(\i\Omega\sigma)\i\Omega\sigma)  \nonumber \\
    &= \sum_{j=0}^\infty \frac{1}{2j + 1}\Tr\left(\left(i\Omega\sigma\right)^{-2j}\right)  \nonumber \\
    &= \sum_{j=0}^\infty \frac{(-1)^j}{2j + 1}\Tr\left(\left(\Omega\sigma\right)^{-2j}\right)  \nonumber \\
    &= \sum_{j=0}^\infty \frac{(-1)^j}{2j + 1}\Tr\left(\left(\Omega(\cosh{(2s)}\mathbb{I} + \sinh{(2s)}M\right))^{-2j}\right)  \nonumber \\
    &= \sum_{j=0}^\infty \frac{(-1)^j}{2j + 1}\Tr\left(\left(\cosh{(2s)}\Omega + \sinh{(2s)}\Omega M\right)^{-2j}\right)  \nonumber \\
    &= 
    \label{eq:A26}
    \sum_{j=0}^\infty \frac{(-1)^j}{2j + 1}\Tr\left(\left(\sinh^2{(2s)} M^2 -\cosh^2{(2s)}\mathbb{I}\right )^{-j}\right).
\end{align}
We use the negative binomial theorem to expand \cref{eq:A26}. For non-negative integers $j$ and real numbers $x$ and $a,$ we have
\begin{align}
    (x+a)^{-n} = \sum_{k=0}^{\infty} (-1)^k \binom{n+k-1}{k}x^k a^{-n-k}
    ,
\end{align}
as discussed in Ref.~\cite{negativebinomial}. This equation can be extended to matrices if $x$ and $a$ commute. Using $x = \sinh^2{(2s)} M^2$ and $a = -\cosh^2{(2s)} \mathbb{I}$, we have
\begin{align}
    S_1^{(2)}
    &= 2k + \sum_{j=1}^\infty \frac{(-1)^j}{2j + 1}\Tr\left(\sum_{i=0}^{\infty} (-1)^i \binom{j+i-1}{i} \sinh^{2i}(2s) M^{2i} (-\cosh^2(2s))^{-j-i}\right) \\
    &= 2k + \sum_{j=1}^\infty \left(\frac{1}{2j + 1} \cdot \sum_{i=0}^{\infty} \binom{j+i-1}{i} \tanh^{2i}(2s) \text{sech}^{2j}(2s) \Tr(M^{2i})\right) \nonumber\\
    &= 2k\cosh{(2s)}\tanh^{-1}{(\text{sech}(2s))} + \sum_{i=1}^{\infty} \tanh^{2i}(2s) \Tr(M^{2i}) \sum_{j=1}^\infty \frac{\text{sech}^{2j}(2s)}{2j + 1} \binom{j+i-1}{i} \nonumber\\
    &= 2k\cosh{(2s)}\tanh^{-1}{(\text{sech}(2s))} + \frac{2}{3} \text{sech}^2(2s) \sum_{i=1}^{\infty} \tanh^{2i} (2s) \, _2F_1\left(\frac{3}{2}, 1+i, \frac{5}{2}, \text{sech}^2 (2s)\right) \Tr (W^i).
\end{align}
Therefore,
\begin{align}
    \begin{split}
        \underset{U \in U(n)}{\mathbb{E}} S_1(U) &= \frac{k}{2} \log \bargs{\frac{\sinh^2 (2s)}{4}} - \sum_{j=1}^\infty \frac{1}{2j} \underset{U}{\mathbb{E}} \Tr W^j\\
        &\quad + k\cosh{(2s)}\tanh^{-1}{(\text{sech}(2s))} + \frac{\text{sech}^2(2s)}{3} \sum_{i=1}^{\infty} \tanh^{2i} (2s) \, _2F_1\left(\frac{3}{2}, 1+i, \frac{5}{2}, \text{sech}^2 (2s)\right) \underset{U}{\mathbb{E}}\Tr (W^i) \\
    \end{split} \\
    \begin{split}
        &= 
        \label{eq:A33}
        nr\left(\frac{1}{2} \log \bargs{\frac{\sinh^2 (2s)}{4}} + \cosh(2s) \tanh^{-1}[\text{sech}(2s)]\right) \\
        &\quad + \frac{\text{sech}^2(2s)}{3} \sum_{i=1}^{\infty}  \left[\tanh^{2i}(2s) \, _2F_1\left(\frac{3}{2}, 1+i, \frac{5}{2}, \text{sech}^2(2s)\right) - \frac{3\cosh^2 (2s)}{2i}\right] \underset{U}{\mathbb{E}}\Tr W^i
    \end{split} \\
    \label{eq:A34}
    &= \sum_{i=1}^{\infty}  \left[\frac{1}{2i} - \frac{1}{3}\textup{sech}^2 (2s) \tanh^{2i}(2s) \, _2F_1\left(\frac{3}{2}, 1+i, \frac{5}{2}, \textup{sech}^2(2s)\right)\right] 
        (nG_{i} (r) - H_{i}(r) + o(1)),
\end{align}
where the step from \cref{eq:A33} to \cref{eq:A34} arises from substituting $\underset{U}{\mathbb{E}}\Tr W^{i} = nr - (nG_{i} (r) - H_{i}(r) + o(1))$ as derived in Ref.~\cite{Iosue_2023} and the subsequent cancellation of the constant term. Finally, dividing by $n$ and noting the vanishing behavior of $H_{i}(r)$ and $o(1)$ as $n \rightarrow \infty,$ the von Neumann Page curve is
\begin{align}
    \label{eq:von-Neumann-Page}
    \begin{split}
        \lim_{n \rightarrow \infty} \underset{U \in U(n)}{\mathbb{E}} \frac{1}{n} S_1(U) &= \sum_{i=1}^{\infty} \left[\frac{1}{2i} - \frac{1}{3}\textup{sech}^2 (2s) \tanh^{2i}(2s) \, _2F_1\left(\frac{3}{2}, 1+i, \frac{5}{2}, \textup{sech}^2(2s)\right)\right] G_{i} (r),
    \end{split}
\end{align}
which is given in \cref{thm:von-Neumann-Page-curve} in the main text.

\subsection{Limit as \texorpdfstring{$s \rightarrow 0$}{s approaches 0}}
\label{ap:S1-zerolimit}

In this section, we show the validity of \cref{thm:von-Neumann-zero}. Moreover, we elucidate the odd behavior of the von Neumann Page curve scaling with $s^2 \log(1/s^2)$ instead of the traditional $s^2$ seen in the other Rényi-$\alpha$ entropies with integer $\alpha \geq 2$. We also give our motivation for investigating the $s^2 \log(1/s^2)$ dependence.

For small squeezing strengths, one might expect the Rényi entanglement entropy to scale with the number of bosons in all modes. Furthermore, since the total expected number of bosons in the state of $n$ modes is $\sum_{i=1}^{n} \sinh^2 {s} \approx ns^2$ for equal squeezing, one may guess that all Rényi-$\alpha$ Page curves scale with $s^2$.

However, 
\begin{align}
    \lim_{s \rightarrow 0} \lim_{n \rightarrow \infty} \underset{U \in U(n)}{\mathbb{E}} \frac{1}{s^2n} S_1(U) = \infty.
\end{align}

One difficulty arises in evaluating the limit as $s \rightarrow \infty$ of the first term in \cref{eq:A33}. Notably, 
\begin{align}
    \label{eq:A59}
   \frac{1}{2s^2} \log \bargs{\frac{\sinh^2 (2s)}{4}} &= \frac{\log s}{s^2} + \frac{2}{3} - \frac{4}{45} s^2 + \frac{64}{2835}s^4 - \frac{32}{4725} s^6 + \frac{1024}{467775}s^8 - \dots, \\
    \label{eq:A60}
   \frac{1}{s^2} \cosh (2s) \tanh^{-1} (\text{sech} (2s)) &= -\frac{\log s}{s^2} + \frac{1-6\log s}{3} + \frac{53-60\log s}{90} s^2 + \frac{251-252\log s}{2835} s^4 + \dots.
\end{align}

Summing Eqs.~\eqref{eq:A59} and \eqref{eq:A60} yields a complicated series involving $\log{s},$ which diverges as $s \rightarrow 0.$ Therefore, we try evaluating $\lim_{s \rightarrow 0} \lim_{n \rightarrow \infty} \underset{U \in U(n)}{\mathbb{E}} \frac{1}{(s^2 \log{1/s^2})n} S_1(U)$ in hopes of achieving a convergent result. Using the entanglement entropy expression derived in \cref{eq:A33},
\begin{align}
    \begin{split}
        \label{eq:A46}
        \lim_{s \rightarrow 0} \lim_{n \rightarrow \infty} \underset{U \in U(n)}{\mathbb{E}} \frac{1}{(s^2 \log{1/s^2}) n} S_1(U) &= \frac{r}{s^2 \log{1/s^2}} \cdot \left(\frac{1}{2} \log \bargs{\frac{\sinh^2 (2s)}{4}} + \cosh(2s) \tanh^{-1}[\text{sech}(2s)]\right) \\
        &\quad + \frac{\text{sech}^2 (2s)}{3s^2 \log(1/s^2)} \sum_{i=1}^{\infty}  \left[\tanh^{2i}(2s) \, _2F_1\left(\frac{3}{2}, 1+i, \frac{5}{2}, \text{sech}^2(2s)\right) - \frac{3\cosh^2 (2s)}{2i}\right] \\
        &\quad \cdot r^{i + 1}C_{i} \, _2F_1 (1-i; i; i + 2; r).
    \end{split}
\end{align}
Note that $\frac{1}{s^2 \log(1/s^2)} \cdot r\left(\frac{1}{2} \log \bargs{\frac{\sinh^2 (2s)}{4}} + \cosh(2s) \tanh^{-1}[\text{sech}(2s)]\right),$ the constant term in \cref{eq:A46}, approaches $r$ as $s \rightarrow 0.$ Now, we need to find the limit as $s \rightarrow 0$ of the second term in \cref{eq:A46}:
\begin{align}
    &= \lim_{s\rightarrow 0} \frac{1}{s^2 \log(1/s^2)} \frac{\text{sech}^2 (2s)}{3} \sum_{i=1}^{\infty}  \left[\tanh^{2i}(2s) \, _2F_1\left(\frac{3}{2}, 1+i, \frac{5}{2}, \text{sech}^2(2s)\right) - \frac{3\cosh^2 (2s)}{2i}\right] r^{i + 1}C_{i} \, _2F_1 (1-i; i; i + 2; r) \nonumber \\
    \label{eq:A48}
    &=  \sum_{i=1}^{\infty}  \lim_{s \rightarrow 0} \frac{1}{s^2 \log(1/s^2)} \left( \frac{1}{3} \text{sech}^2 (2s) \tanh^{2i}(2s) \, _2F_1\left(\frac{3}{2}, 1+i, \frac{5}{2}, \text{sech}^2(2s)\right) - \frac{1}{2i} \right)\cdot r^{i + 1}C_{i} \, _2F_1 (1-i; i; i + 2; r) \nonumber \\
    &= \sum_{i=1}^{\infty}  \lim_{s \rightarrow 0} \frac{1}{s^2 \log(1/s^2)} \left(\frac{1}{3} \cdot 1\cdot (2s)^{2i} \, _2F_1\left(\frac{3}{2}, 1+i, \frac{5}{2}, (1-2s^2)^2\right) - \frac{1}{2i} \right)\cdot r^{i + 1}C_{i} \, _2F_1 (1-i; i; i + 2; r).
\end{align}

Note that, in the $i=1$ case, letting $F(s) = \frac{1}{s^2 \log(1/s^2)} \frac{1}{3} \cdot 1\cdot (2s)^{2i},$ the left-hand derivative $\frac{dF}{ds_{s = 0^{-}}} = -\infty$, while the right-hand derivative $\frac{dF}{ds_{s = 0^{+}}} = \infty.$ This behavior deviates from all other values of $i \geq 1,$ where $\frac{dF}{ds} = 0$ evaluated at $s = 0.$ Combined with the dependence of $_2F_1\left(\frac{3}{2}, 1+i, \frac{5}{2}, (1-2s^2)^2\right)$ on $i,$ one can explicitly calculate that \cref{eq:A48} evaluates to $-r^2,$ as the $i=1$ case yields $-r^2$ and subsequent $i$ yield $0$ in the sum. Thus, 
\begin{align}
    \lim_{s \rightarrow 0} \lim_{n \rightarrow \infty} \underset{U \in U(n)}{\mathbb{E}} \frac{1}{(s^2 \log(1/s^2))n} S_1(U) &= r(1-r).
\end{align}
It can easily be seen from \cref{eq:A33} that the result is the same if we swap the limits, so that by the Moore-Osgood theorem \cite{taylor2012general},
\begin{align}
    \label{eq:von-Neumann-zero}
    \lim_{\substack{n\to\infty \\ s \rightarrow 0 }} \underset{U \in U(n)}{\mathbb{E}} \frac{1}{(s^2 \log(1/s^2))n} S_1(U) &= r(1-r).
\end{align}
as displayed in \cref{thm:von-Neumann-zero} in the main text.

\subsection{Limit as \texorpdfstring{$s \rightarrow \infty$}{s approaches infinity}}
\label{ap:S1-infinitylimit}
In a similar vein as \cref{ap:S1-zerolimit}, we also compute the von Neumann Page curve for equal squeezing as $s\rightarrow \infty.$ The Page curve scales with $s$ in the limit $s\to \infty$, so we compute $\lim_{s \rightarrow \infty} \lim_{n \rightarrow \infty} \underset{U \in U(n)}{\mathbb{E}} \frac{1}{sn} S_1(U).$ We once again use the expression found in \cref{eq:A33} in our evaluations.
\begin{align}
    \begin{split}
        \lim_{s \rightarrow \infty} \lim_{n \rightarrow \infty} \underset{U \in U(n)}{\mathbb{E}} \frac{1}{sn} S_1(U) &= \lim_{s \rightarrow \infty} A_1 + \lim_{s \rightarrow \infty} A_2 ,\\
        \text{where } A_1 &= \frac{r}{s} \left(\frac{1}{2}\log \bargs{\frac{\sinh^2 (2s)}{4}} + \cosh(2s) \tanh^{-1}[\text{sech}(2s)]\right) ,\\
        A_2 &= \frac{1}{s} \sum_{i=1}^{\infty} \left[\frac{1}{3} \text{sech}^2 (2s) \tanh^{2i}(2s) \, _2F_1\left(\frac{3}{2}, 1+i, \frac{5}{2}, \text{sech}^2(2s)\right) - \frac{1}{2i}\right]\cdot f_i(r), \\
        f_i(r) &= r^{i + 1}C_{i} \, _2F_1 (1-i; i; i + 2; r).
    \end{split}
\end{align}
It turns out that
\begin{align}
    \lim_{s\rightarrow \infty} A_1 &= \lim_{s \rightarrow \infty} \left(\frac{r}{s} \frac{1}{2}\log \bargs{\frac{\sinh^2 (2s)}{4}}\right) + \lim_{s \rightarrow \infty} \left(\frac{r}{s}\cdot \cosh(2s) \tanh^{-1}[\text{sech}(2s)]\right) \\
    &= \frac{r}{2} \lim_{s\rightarrow \infty} \left(\frac{1}{s} \log \bargs{\frac{\sinh^2 (2s)}{4}} \right) + 0 \nonumber \\
    &= 2\min(r, 1-r). \label{eq:a27}
\end{align}

Since the Page curve is symmetric about $r = \frac{1}{2},$ \cref{eq:a27} contains a $\min(r, 1-r)$ term (as opposed to a simple $r$-dependence). Next, we calculate $\lim_{s\rightarrow \infty} A_2$:
\begin{align}
    \lim_{s\rightarrow \infty} A_2 &= \lim_{s\rightarrow \infty} \frac{1}{s} \sum_{i=1}^{\infty} \left[\frac{1}{3} \text{sech}^2 (2s) \tanh^{2i}(2s) \, _2F_1\left(\frac{3}{2}, 1+i, \frac{5}{2}, \text{sech}^2(2s)\right) - \frac{1}{2i}\right]\cdot f_i(r) \\
    &= \lim_{s\rightarrow \infty} \frac{1}{s} \sum_{i=1}^{\infty} \left[\frac{1}{3} \cdot 0\cdot 1 \, _2F_1\left(\frac{3}{2}, 1+i, \frac{5}{2}, 0\right) - \frac{1}{2i}\right]\cdot f_i(r) \\
    &= 0.
\end{align}
Therefore,
\begin{align}
    \label{eq:von-Neumann-inf}
    \lim_{\substack{n\to\infty \\s \rightarrow \infty}} \underset{U \in U(n)}{\mathbb{E}} \frac{1}{sn} S_1(U) &= 2\min(r, 1-r),
\end{align}
where we again used the same discussion as around \cref{eq:von-Neumann-zero}

\cref{eq:von-Neumann-inf} is verified in \cref{fig:inf}, alongside other Rényi-$\alpha$ Page curves in the limit as $s\rightarrow \infty.$ Note that both the von Neumann and Rényi Page curves approach the same limit of $2 \min(r, 1-r)$ (see \cref{ap:Salpha-infinity} for the relevant derivation). This indicates that, at large squeezing values, the von Neumann entropy behaves similarly to the larger class of Rényi entropies.

\section{General \texorpdfstring{Rényi-$\alpha$}{Rényi-alpha} Entropies}
\label{ap:Salpha}

In this section, we extend our derivation in \cref{ap:S1-appendix} from the von Neumann entropy to all general Rényi-$\alpha$ entropies, where $\alpha$ is a positive integer greater than $1$. We find a closed formula for the Rényi-$\alpha$ Page curve as a function of equal squeezing strength $s$ and partition ratio $r$ in the asymptotic limit $n \rightarrow \infty.$ Similar to the von Neumann case, we calculate squeezing limits of the Rényi-$\alpha$ entropies as $s\rightarrow 0$ and $s\rightarrow \infty$ and then discuss typicality of entanglement based on the variance.

\subsection{Page Curve for the \texorpdfstring{Rényi-$\alpha$}{Rényi-alpha} entropy}
\label{ap:Salpha-pagecurve}
Here, we present our full derivation for the Page curve of the Rényi-$\alpha$ entropy, where $\alpha$ is an integer great than 1, in terms of the partition ratio $r$ and equal squeezing strength $s$.

Our starting point is the general expression relating the Rényi-$\alpha$ entropy with the symplectic eigenvalues $\nu_j$ of the covariance matrix $\sigma$ (recall that $\sigma$ is precisely the second moment of $\rho$ \cite{serafini_quantum_2017}):
\begin{align}
& \mathcal{S} _{\alpha} (\rho )
= \frac{1}{\alpha-1}\sum_{j=1}^k  \log\bargs{(\nu_j+1)^\alpha - (\nu_j - 1)^\alpha} - \frac{\alpha k \log 2 }{\alpha-1} 
.
\end{align}
If $\alpha$ is a positive integer, then this function has $\alpha-1$ singularities. We can solve for those singularities explicitly:
\begin{align}
\label{neq:Salpha_zeros_sol}
&(\nu _{0}+1) ^\alpha
= (\nu _{0}- 1) ^\alpha 
\Rightarrow 
\nu _{0}+1 = 
(\nu _{0}- 1) e ^{i \frac{ 2\pi \ell}{\alpha}}
, \\
& \Rightarrow 
\nu _{0}= - i \cot  
\frac{\pi \ell }{\alpha}
, \quad
\ell = 1, \ldots, \alpha-1
.
\end{align}
Note that we exclude the $\ell = \alpha$ case as Eq.~\eqref{neq:Salpha_zeros_sol} does not have a root in that case. Thus, the polynomial $(\nu _{j}+1) ^\alpha
- (\nu _{j}- 1) ^\alpha$ can be rewritten in terms of a product of factors as follows:
\begin{align}
\label{eq:B4}
(\nu _{j}+1) ^\alpha
- (\nu _{j}- 1) ^\alpha
= 2 \alpha 
\nu _{j} ^{\zeta} \prod 
\limits_{m=1} ^{ \lfloor 
\frac{\alpha -1}{2} \rfloor } 
(\nu _{j}^{2} + \cot ^{2} 
\frac{\pi m }{\alpha} )
, 
\end{align}
where we define $\zeta := 1-(\alpha\bmod 2)$. 
The R{\'{e}}nyi entropy is in turn given by 
\begin{align}
\label{neq:Salpha_zeros}
& \mathcal{S} _{\alpha} (\rho )
= \frac{1}{\alpha -1 }
\sum _{m=1} ^{ \lfloor 
\frac{\alpha -1}{2} \rfloor }
\sum _{j=1} ^{k} 
\log 
( \nu _{j}^{2} + \cot ^{2} 
\frac{\pi m }{\alpha} ) 
+ \frac{\zeta \sum _{j=1} ^{k}  \log \nu _{j}^{2} }{2(\alpha -1)}
+ \frac{ k\log \alpha }
{\alpha -1 } - k \log 2
.
\end{align}
To work with the $\log \nu_j^2$ terms, note that the squares of the symplectic eigenvalues of $\sigma$ are precisely the eigenvalues of $(i\Omega \sigma)^2$ by matrix properties. Furthermore, since the eigenvalues of $i\Omega \sigma$ come in plus-minus pairs but the symplectic eigenvalues only consider positive numbers, the factor of $\frac{1}{2}$ in \cref{neq:Salpha_zeros} arises. Taking $(i \Omega \sigma ) ^{2} = 
(i \Omega (\mathbb{I}\cosh (2s) 
+ M \sinh (2s) ) ) ^{2}
= \cosh^2 (2s) 
- M^{2} \sinh^2 (2s) $ into the above equation, we thus have 
\begin{align}
\mathcal{S} _{\alpha} (\rho )
= &  \frac{1}{ 2 (\alpha -1) }
\sum _{m=1} ^{ \lfloor 
\frac{\alpha -1}{2} \rfloor }
\text{Tr}
\log 
( \cot ^{2} 
\frac{\pi m }{\alpha} 
+ \cosh^2 (2s) 
- M^{2} \sinh^2 (2s))
\nonumber \\
\label{neq:Salpha_int}
&+ \frac{\zeta }{4(\alpha -1)}
\text{Tr}
\log 
(  \cosh^2 (2s)
- M^{2} \sinh^2 (2s))
+ \frac{ k\log ( \alpha) }
{\alpha -1 } - k \log 2
.
\end{align}

Let us now expand all the logarithmic functions in the sum in Eq.~\eqref{neq:Salpha_int} in orders of $M^{2}$. Noting that $M^{2} \le 1$, one could show that each of the following series always converge (for any integer $\ell$): 
\begin{align}
\label{eq:B7}
\text{Tr} \log 
(  \cosh^2 (2s)
- M^{2} \sinh^2 (2s) ) &= 4 k\log \cosh (2s)
- \sum _{ \ell=1}^{\infty }  \frac{\sinh^{2  \ell} (2s) }
    { \ell \cosh^{2  \ell} (2s) }
    \Tr (M ^{2 \ell})
,\\
\label{eq:B8}
\text{Tr} \log 
( \cot ^{2} \frac{\pi m }{ \alpha} 
+ \cosh^2 (2s)
- M^{2} \sinh^2 (2s) ) &= 2 k\log 
( \cot ^{2} 
\frac{\pi m }{\alpha} 
+ \cosh^2 (2s)) 
- \sum _{ \ell=1}^{\infty }  \frac{ \sinh^{2 \ell}(2s) }
    { \ell (\cosh^2(2s) + \cot ^{2} 
\frac{\pi m }{\alpha} )^{ \ell}}
    \Tr (M ^{2 \ell})
.
\end{align}
Note that the product in \cref{eq:B4} vanishes when substituting $\alpha = 2$. Therefore, one gets from \cref{neq:Salpha_zeros} that $S_2 = \sum_{j=1}^{k} \log \nu_j^2$. Combining the expansion in \cref{eq:B7}, we get that $S_2 = k \log \cosh(2s) - \sum_{\ell = 1}^{\infty} \frac{\tanh^{2\ell} (2s)}{2\ell} \Tr (M^{2\ell}).$ This expression for $S_2$ is utilized in the following steps to formally express the R{\'{e}}nyi entropy with a positive integer $\alpha$ as
\begin{align}
\mathcal{S} _{\alpha} (\rho )
= & k \sum _{m=1} ^{ \lfloor 
\frac{\alpha -1}{2} \rfloor }
\frac{\log 
( \cot ^{2} 
\frac{\pi m }{\alpha} 
+ \cosh^2 (2s) ) }{\alpha -1 }
+ k \frac{ \zeta \log \cosh (2s)
+ \log \alpha }
{\alpha -1 }
- k \log 2
\\
& - \frac{1}{\alpha -1 }
\sum _{\ell=1}^{\infty } \left [
\sum _{m=1} ^{ \lfloor 
\frac{\alpha -1}{2} \rfloor }
\frac{ \sinh^{2\ell}(2s) }
    { 
    \ell (\cosh^2(2s) + \cot ^{2}
\frac{\pi m }{2\alpha} )^{ \ell}} 
+ \zeta \frac{\sinh^{2\ell} (2s) }
{2 \ell  \cosh^{2\ell} (2s) } \right ]
    \frac{ \Tr (M ^{2\ell})}{2 } \nonumber
\\
\label{eq:B13}
= & k \sum _{m=1} ^{ \lfloor 
\frac{\alpha -1}{2} \rfloor }
\frac{\log 
( \cot ^{2} 
\frac{\pi m }{\alpha} 
+ \cosh^2 (2s) ) }{\alpha -1 }
+ \frac{ k\log \alpha}
{\alpha -1 } - k \log 2+ \frac{ \zeta } 
{\alpha -1 } S_2 \nonumber
\\
& - \frac{1}{\alpha -1 }
\sum _{\ell=1}^{\infty } \left [
\sum _{m=1} ^{ \lfloor 
\frac{\alpha -1}{2} \rfloor }
\frac{ \sinh^{2\ell}(2s) }
    {
\ell (\cosh^2(2s) + \cot ^{2}
\frac{\pi m }{\alpha} )^{\ell}}  
\right ]
   \frac{ \Tr (M ^{2\ell})}{2 }
\\
\label{eq:B14}
= &  \frac{ \zeta }  {\alpha -1 } S_2
+ \frac{1}{\alpha -1 }
\sum _{\ell=1}^{\infty } \left [
\sum _{m=1} ^{ a }
\frac{ \sinh^{2\ell}(2s) }
    {
\ell (\cosh^2(2s) + \cot ^{2}
\frac{\pi m }{\alpha} )^{\ell}}  
\right ]
\left [ k- \Tr (W ^{\ell})
\right ]
    , 
\end{align}
where $a \equiv  \lfloor \frac{\alpha -1}{2} \rfloor 
= \frac{\alpha -1 - \zeta }{2}$. We have used $\Tr M ^{2\ell} = 2 \Tr W ^{\ell} $ in \cref{eq:B14}. The step from \cref{eq:B13} to \cref{eq:B14} utilizes the relation presented in \cref{eq:B8}, where $M$ is substituted with the identity matrix. We can then use the expression for $\bbE_{U\in\U(n)} \Tr W^\ell$ from Ref.~\cite{Iosue_2023} to derive the Page curve.

\subsection{Limit as \texorpdfstring{$s\rightarrow 0$}{s approaches 0}}
\label{ap:Salpha-zerolimit}
We now analyze the Rényi-$\alpha$ entropy as the squeezing strength approaches zero. This time, due to the analyticity of the Rényi entropies for integer $\alpha \geq 2$, the Page curve scales with the number of bosons $\sim s^2$ instead of the $s^2 \log(1/s^2)$ dependence explained in \cref{ap:S1-zerolimit}.

\begin{figure}[H]
    \centering
    \includegraphics[width=.8\textwidth]{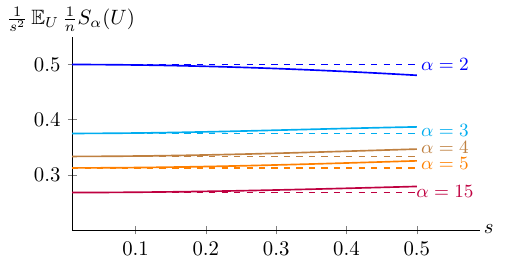}
    \caption{The Rényi-$\alpha$ Page curves, for $\alpha \in \{2, 3, 4, 5, 15\},$ divided by $s^2$, as a function of the squeezing strength $s.$ These functions are displayed as the solid lines. The Rényi-$\alpha$ Page curve limit as $s\rightarrow 0,$ presented in \cref{eq:B21}, is shown by the dashed lines in corresponding color. As expected, the Rényi-$\alpha$ Page curve approaches $\frac{\alpha}{\alpha - 1}r(1-r)$ in order $s^2,$ where we take $r=0.5$ and $n=400$.}
    \label{fig:zero}
\end{figure}

Recall that $G_{\ell} (r)$ is a function symmetric about $r = \frac{1}{2}$ defined as $r - r^{\ell+1} C_{\ell} \, _2F_1(1-\ell; \ell; \ell + 2; r)$. Using \cref{eq:B14} and substituting $k - \Tr (W^{\ell}) = nG_{\ell}(r)$ to investigate the $s\rightarrow 0$ limit,
\begin{align}
    \lim_{s\to 0}\lim_{n\to\infty}\frac{1}{s^2 n}\Expval_U S_\alpha(U) &= \lim_{s\to 0} \frac{1}{s^2} \frac{1}{\alpha -1} \sum _{\ell=1}^\infty \frac{1}{\ell} (2 s)^{2 \ell} G_\ell(r) \sum _{m=1}^a \left(1 + \cot ^2\left(\frac{\pi  m}{\alpha }\right)\right)^{-\ell}  +\frac{\zeta }{\alpha - 1} 2 r(1-r) \\
    &= \lim_{s\to 0} \frac{1}{s^2} \frac{1}{\alpha -1} (2 s)^{2} G_1(r) \sum _{m=1}^a \left(\cot ^2\left(\frac{\pi  m}{\alpha }\right)+1\right)^{-1} +\frac{\zeta }{\alpha -1} 2 r(1-r) \nonumber\\
&= \frac{4}{\alpha -1} r(1-r) \sum _{m=1}^a \left(\cot ^2\left(\frac{\pi  m}{\alpha }\right)+1\right)^{-1} +\frac{\zeta }{\alpha -1} 2 r(1-r) \nonumber\\
&= \frac{r(1-r)}{\alpha-1} \parentheses{4 \sum _{m=1}^a \left(\cot ^2\left(\frac{\pi  m}{\alpha }\right)+1\right)^{-1} + 2\zeta} \nonumber \\
&= \frac{r(1-r)}{\alpha-1} \parentheses{\left(-\csc \left(\frac{\pi }{\alpha }\right) \sin \left(\frac{2 \pi  a+\pi }{\alpha }\right)+2 a+1\right) + 2\zeta}.
\end{align}
Next, we use that $a = \floor{(\alpha - 1)/2} = (\alpha - 1 - \zeta)/2$. Then
\begin{align}
    \lim_{s\to 0}\lim_{n\to\infty}\frac{1}{s^2 n}\Expval_U S_\alpha(U) &= \frac{r(1-r)}{\alpha - 1}\parentheses{ \alpha + \zeta - \frac{\sin(\pi \zeta/\alpha)}{\sin(\pi/\alpha)} } \\
    \label{eq:B21}
    &= \frac{\alpha}{\alpha - 1} \cdot r (1-r).
\end{align}
As in the other cases, this results holds as a double limit $\lim_{\substack{n\to\infty \\ s\to 0}}$.
Evidently, the Rényi-$\alpha$ Page Curve with small squeezing approaches $r(1-r)\cdot s^2$ as $\alpha \rightarrow \infty.$ 
We demonstrate \cref{eq:B21} numerically in \cref{fig:zero}.

\subsection{Limit as \texorpdfstring{$s\rightarrow \infty$}{s approaches infinity}}
\label{ap:Salpha-infinity}
Now, we investigate the squeezing limit $s\rightarrow \infty.$

We have 
\begin{align}
\begin{split}
    \lim_{s\to \infty}\lim_{n\to\infty}\frac{1}{s n}\Expval_U S_\alpha(U) &= \lim_{s\to \infty} \frac{1}{s} \frac{1}{\alpha -1} \sum _{\ell=1}^\infty \frac{1}{\ell} \sinh^{2 \ell} (2s) G_\ell(r) \sum _{m=1}^a \left(\cot ^2\left(\frac{\pi  m}{\alpha }\right)+\cosh^2 (2s) \right)^{-\ell} \\
    &\qquad +\frac{\zeta }{\alpha -1} 2 \text{ min} (r, 1-r)
\end{split}\\
&= \lim_{s\to \infty} \frac{1}{s} \frac{2a}{\alpha -1} \left(\sum _{\ell=1}^\infty \frac{\tanh^{2 \ell} (2s)}{2\ell}  G_\ell(r)\right) +\frac{\zeta }{\alpha -1} 2 \text{ min} (r, 1-r) \nonumber \\
&= \frac{2a}{\alpha-1} 2 \text{ min} (r, 1-r) + \frac{\zeta}{\alpha - 1} 2 \text{ min} (r, 1-r) \nonumber \\
&= \frac{4a + 2\zeta}{\alpha - 1} \text{ min} (r, 1-r) \nonumber \\
\label{eq:B26}
&= 2\text{ min} (r, 1-r).
\end{align}
As in the other cases, this results holds as a double limit $\lim_{\substack{n\to\infty \\ s\to 0}}$.

We numerically validate \cref{eq:B26} and \cref{eq:von-Neumann-inf} in \cref{fig:inf} below. As shown, both the von Neumann and Rényi-$\alpha$ entropies indeed approach $2 \min (r, 1-r)$ in order $s$ as $s\rightarrow \infty.$ 

\begin{figure}[H]
    \centering
    \includegraphics[width=.8\textwidth]{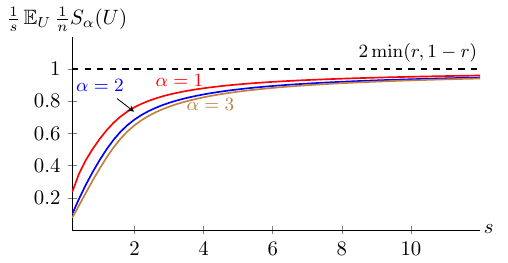}
    \caption{Plots of the von Neumann, Rényi-2, and Rényi-3 Page curves divided by $s$, as a function of the squeezing strength $s.$ These functions, which are presented more generally in \cref{thm:von-Neumann-inf} and \cref{thm:Renyi-zero-inf}, are shown by the solid red, blue, and brown lines, respectively. The von Neumann and Rényi Page curves in the limit as squeezing strength approaches infinity, derived in \cref{ap:S1-infinitylimit} and \cref{ap:Salpha-infinity}, have coinciding values of $2\min(r, 1-r)$ displayed as the black dotted line. We take $r=0.5$.}
    \label{fig:inf}
\end{figure}

\subsection{Variance and typicality of entanglement}
\label{ap:Salpha-entanglement}

In this section, we prove weak typicality of entanglement described in \cref{sssec:Salpha-entanglement}. We consider the general expression for the Rényi-$\alpha$ entropy as found in \cref{eq:B14}:
\begin{align}
\mathcal{S} _{\alpha} (\rho )
&= \frac{ \zeta }  {\alpha -1 } S_2
+ \frac{1}{\alpha -1 }
\sum _{\ell=1}^{\infty } \left [
\sum _{m=1} ^{ a }
\frac{ \sinh^{2\ell}(2s) }
    {
\ell (\cosh^2(2s) + \cot ^{2}
\frac{\pi m }{\alpha} )^{\ell}}  
\right ]
\left [ k- \Tr (W ^{\ell})
\right ]
\\
&= \frac{1}{\alpha -1 }
\sum _{\ell=1}^{\infty } \left [
\sum _{m=1} ^{ \lfloor 
\frac{\alpha -1}{2} \rfloor }
\frac{ \sinh^{2\ell}(2s) }
    { 
    \ell (\cosh^2(2s) + \tan ^{2}
\frac{\pi m }{2\alpha} )^{ \ell}} 
+ \zeta \frac{\sinh^{2\ell} (2s) }
{2 \ell  \cosh^{2\ell} (2s) } \right ]
\left [ k- \Tr (W ^{\ell})
\right ] 
, 
\end{align}
with the same definition $a :=  \lfloor \frac{\alpha -1}{2} \rfloor 
= \frac{\alpha -1 - \zeta }{2}$. More specifically, the $\alpha = 2$ R{\'{e}}nyi entropy is given by 
\begin{align}
& \mathcal{S} _{\alpha =2 } (\rho ) = \sum _{\ell=1}^{\infty } 
\frac{\sinh^{2\ell} (2s) }
{2 \ell  \cosh^{2\ell} (2s) } 
\left [ k- \Tr (W ^{\ell})
\right ] 
. 
\end{align}
From Ref.~\cite{Iosue_2023}, we know that the following covariance is a non-zero constant independent of $n$:
\begin{align}
\mathcal{V} _{d} \equiv 
\sum _{ \ell =1} ^{d-1} \frac{1}{ \ell (d-\ell)}
( \overline{ \Tr (W ^{\ell}) \Tr (W ^{d-\ell}) } 
- \overline{ \Tr (W ^{\ell}) } \cdot \overline{ \Tr (W ^{d-\ell}) } ).
\end{align}
Then the variance of 
$\mathcal{S} _{\alpha =2 } (\rho )$ can also be proved to be a constant. Letting the fluctuation in the R{\'{e}}nyi entropy be defined as 
$\delta \mathcal{S} _{\alpha =2 } 
= \mathcal{S} _{\alpha =2 } (\rho ) 
- \overline{\mathcal{S} _{\alpha =2 } (\rho )}$, we get $\text{Var} \, S_{\alpha = 2}(\rho) = \overline{ \delta \mathcal{S} _{\alpha =2 } ^{2} } $ as
\begin{align}
\overline{ \delta \mathcal{S} _{\alpha =2 } ^{2} } 
& = \frac{1}{4}\sum _{ d=2 }^{\infty} \tanh^{2d} (2s) 
\sum _{ \ell =1} ^{d-1} \frac{1}{ \ell (d-\ell)}
( \overline{ \Tr (W ^{\ell}) \Tr (W ^{d-\ell}) } 
- \overline{ \Tr (W ^{\ell}) } \cdot \overline{ \Tr (W ^{d-\ell}) } )
\\
& = \frac{1}{4}\sum _{ d=2 }^{\infty} \tanh^{2d} (2s) \mathcal{V} _{d} 
. 
\end{align}

On the other hand, we can also expand the R{\'{e}}nyi entropy as 
\begin{align}
\mathcal{S} _{\alpha} (\rho )
&= \frac{1}{\alpha -1 } \sum _{m=1} ^{ \lfloor 
\frac{\alpha -1}{2} \rfloor } \mathcal{C} _{m } (\alpha) + 
\frac{\zeta}{\alpha -1 } \mathcal{S} _{2} (\rho )
\\
&= \frac{1}{\alpha -1 }
\left [
\sum _{m=1} ^{ \lfloor 
\frac{\alpha -1}{2} \rfloor }
\sum _{\ell=1}^{\infty } \left ( \frac{ \sinh^{2 }(2s) } {\cosh^2(2s) + \cot ^{2}
( \frac{\pi m }{2\alpha}) } \right )^{ \ell} 
 \frac{\left [ k- \Tr (W ^{\ell})
\right ] }{\ell }
+ \frac{\zeta }{2 } \sum _{\ell=1}^{\infty }\tanh^{2\ell} (2s)  \frac{\left [ k- \Tr (W ^{\ell})
\right ] }{\ell }
\right ]
, 
\end{align}
where
\begin{align}
\mathcal{C} _{m } (\alpha) = \sum _{\ell=1}^{\infty } \left ( \frac{ \sinh^{2 }(2s) } {\cosh^2(2s) + \cot ^{2}
( \frac{\pi m }{2\alpha}) } \right )^{ \ell} 
 \frac{\left [ k- \Tr (W ^{\ell})
\right ] }{\ell }
, \quad 
m = 1,2, \ldots, \floor{ \frac{\alpha -1}{2} } .  
\end{align}
Note that all quantities $\mathcal{C} _{m } (\alpha) $ are well-defined and converge for all integers $m$, $\alpha$, and $s$, and, by definition, $\mathcal{C} _{0 } (\alpha) = \mathcal{S} _{\alpha =2 } (\rho ) $. Further, for each $\mathcal{C} _{m } (\alpha) $, if we introduce its fluctuation 
$ \delta \mathcal{C} _{m } (\alpha) = \mathcal{C} _{m } (\alpha) - \overline{\mathcal{C} _{m } (\alpha) } $ , then 
\begin{align}
\overline{ \delta \mathcal{C} ^{2} _{m } (\alpha) } 
& = \frac{1}{4}\sum _{ d=2 }^{\infty} \left ( \frac{ \sinh^{2 }(2s) } {\cosh^2(2s) + \cot ^{2}
( \frac{\pi m }{2\alpha}) } \right ) ^{d} 
\sum _{ \ell =1} ^{d-1} \frac{1}{ \ell (d-\ell)}
( \overline{ \Tr (W ^{\ell}) \Tr (W ^{d-\ell}) } 
- \overline{ \Tr (W ^{\ell}) } \cdot \overline{ \Tr (W ^{d-\ell}) } )
\\
& = \frac{1}{4}\sum _{ d=2 }^{\infty} \left ( \frac{ \sinh^{2 }(2s) } {\cosh^2(2s) + \cot ^{2}
( \frac{\pi m }{2\alpha}) } \right ) ^{d} \mathcal{V} _{d} \\ \nonumber
&\leq \frac{1}{4} \sum_{d=2}^{\infty} \tanh^{2d} (2s) \mathcal{V}_d,
\end{align}
meaning that each $C_m$ has smaller variance than that of $C_0$. Furthermore, by the Cauchy-Schwarz inequality, we have that 
\begin{align}
\overline{ \delta \mathcal{C}  _{m } (\alpha) \delta \mathcal{C}  _{m' } (\alpha) } 
\le \sqrt{\overline{ \delta \mathcal{C}  _{m } (\alpha) \delta \mathcal{C}  _{m } (\alpha) } 
\cdot 
\overline{ \delta \mathcal{C}  _{m' } (\alpha) \delta \mathcal{C}  _{m' } (\alpha) } }
 \quad \forall m , m' \in {0,1, \ldots, \floor{ \frac{\alpha -1}{2} }},
\end{align}
so that covariances of the form $\overline{ \delta \mathcal{C}  _{m } (\alpha) \delta \mathcal{C}  _{m' } (\alpha) }$  $\forall m , m'$ must also be bounded by a constant as $n$ increases. Finally, since we have found the variance of $\mathcal{C}_{m}(\alpha)$ and $S_2 (\rho)$ to be constant in $n,$ we can prove the concentration of R{\'{e}}nyi entropy with integer $\alpha \ge 2$:
\begin{align}
& \mathcal{S} _{\alpha} (\rho )
= \frac{1}{\alpha -1 } \sum _{m=1} ^{ \lfloor 
\frac{\alpha -1}{2} \rfloor } \mathcal{C} _{m } (\alpha) + 
\frac{\zeta}{\alpha -1 } \mathcal{S} _{2} (\rho )
\Rightarrow 
\overline{ \delta \mathcal{S} _{\alpha } ^{2} } 
= \text{const.}
\end{align}

\twocolumngrid
\bibliography{references}

\end{document}